\documentclass{JHEP3}
\usepackage{graphics}
\newcommand{\re}{\mathop{\mathrm{Re}}\nolimits}
\newcommand{\im}{\mathop{\mathrm{Im}}\nolimits}
\preprint{PITHA-07/04, SLAC-PUB-12567}
\title{On one master integral
for three-loop on-shell HQET propagator diagrams with mass}

\author{A.G.~Grozin\\
Budker Institute of Nuclear Physics, Novosibirsk, Russia, and\\
Department of Physics, University of Alberta, Edmonton, Canada\\
\email{A.G.Grozin@inp.nsk.su}}

\author{T.~Huber\\
Institut f\"{u}r Theoretische Physik E, RWTH Aachen,
D-52056 Aachen, Germany\\
\email{thuber@physik.rwth-aachen.de}}

\author{D.~Ma\^{\i}tre\\
Stanford Linear Accelerator Center, Stanford University,
Stanford, CA 94309, USA\\
\email{maitreda@slac.stanford.edu}}

\abstract{An exact expression for the master integral $I_2$~\cite{GSS:06}
arising in three-loop on-shell HQET propagator diagrams with mass
is derived and its analytical expansion
in the dimensional regularization parameter $\varepsilon$ is given.}


\begin{document}

\FIGURE{
\begin{picture}(50,28.5)
\put(25,14.25){\makebox(0,0){\includegraphics{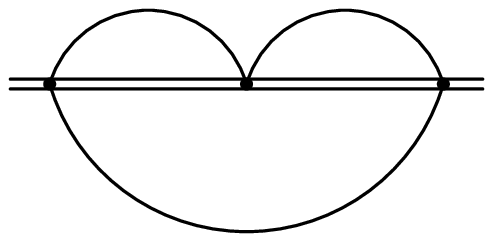}}}
\put(25,0){\makebox(0,0)[b]{$n_3$}}
\put(15,28.5){\makebox(0,0)[t]{$n_2$}}
\put(35,28.5){\makebox(0,0)[t]{$n_2$}}
\put(15,16.5){\makebox(0,0)[t]{$n_1$}}
\put(35,16.5){\makebox(0,0)[t]{$n_1$}}
\end{picture}
\caption{The integral $I_{n_1 n_2 n_3}$}
\label{Diag}}

All three-loop on-shell HQET propagator integrals
with a loop of a massive quark can be reduced
to a set of master integrals~\cite{GSS:06}.
A reduction algorithm has been constructed
by solving the integration-by-parts~\cite{CT:81} relations
using a Gr\"obner bases technique~\cite{SS:06}.
Some master integrals are known exactly,
for the remaining ones a few terms of their respective
$\varepsilon$ expansion
have been calculated in Ref.~\cite{GSS:06}.
In particular, for the integral $I_2$,
the $1/\varepsilon$ and $\mathcal{O}(\varepsilon^0)$ terms are known.
However, in some applications
more terms of its $\varepsilon$ expansion are needed
(A.~Czarnecki, A.~Pak, work in progress).
Here, we shall obtain an exact result for $I_2$
and additional terms of its expansion in $\varepsilon$.

We shall consider a more general integral (Fig.~\ref{Diag}):
\begin{eqnarray}
I_{n_1 n_2 n_3} &=& \frac{1}{i\pi^{d/2}}
\int \frac{I_{n_1 n_2}^2(p_0)\,d^d p}{(1-p^2-i0)^{n_3}}\,,
\nonumber\\
I_{n_1 n_2}(p_0) &=& \frac{1}{i\pi^{d/2}}
\int \frac{d^d k}{(-2(k_0+p_0)-i0)^{n_1} (1-k^2-i0)^{n_2}}\,,
\label{I123}
\end{eqnarray}
using a method similar to the one used in~\cite{B:92}
for calculating a three-loop vacuum integral.
Using the HQET Feynman parametrization,
we write the one-loop subdiagram as
\begin{equation}
I_{n_1 n_2}(p_0) =
\frac{\Gamma(n_1+n_2-2+\varepsilon)}{\Gamma(n_1) \Gamma(n_2)}
\int_0^\infty y^{n_1-1} (y^2 - 2 p_0 y + 1)^{2-n_1-n_2-\varepsilon}\,dy\,.
\label{Fey}
\end{equation}
Then, after the Wick rotation $p_0=i p_{E0}$,
we can calculate the integral in $d^{d-1}\vec{p}$:
\begin{equation}
I_{n_1 n_2 n_3} =
\frac{\Gamma(n_3-3/2+\varepsilon)}{\pi^{1/2}\Gamma(n_3)}
\int_{-\infty}^{+\infty}
I_{n_1 n_2}^2(i p_{E0}) (1+p_{E0}^2)^{3/2-n_3-\varepsilon} d p_{E0}\,.
\label{Wick}
\end{equation}
Here $\re I_{n_1 n_2}^2(i p_{E0})$ is an even function of $p_{E0}$;
$\im I_{n_1 n_2}^2(i p_{E0})$ is an odd function,
and does not contribute to the integral.

The integral~(\ref{Fey}) at $p_0<0$
can be expressed via the hypergeometric function ${}_2\!F_1$:
\begin{equation}
I_{n_1 n_2}(p_0) =
\frac{\Gamma(n_1+n_2-2+\varepsilon) \Gamma(n_1+2n_2-4+2\varepsilon)}%
{\Gamma(n_2) \Gamma(2(n_1+n_2-2+\varepsilon))}
{}_2\!F_1 \left( \left.
\begin{array}{c}
\frac{1}{2} n_1, \frac{1}{2} n_1 + n_2 - 2 + \varepsilon\\
n_1 + n_2 - \frac{3}{2} + \varepsilon
\end{array}
\right| 1 - p_0^2 \right)
\label{I12}
\end{equation}
(a similar expression has been derived in Ref.~\cite{Z:02}).
Its square can be expressed via ${}_3\!F_2$,
due to the Clausen identity.
In order to perform an analytical continuation to $p_0=i p_{E0}$,
we re-express this ${}_3\!F_2$ via three ${}_3\!F_2$ functions
of the inverse argument:
\begin{equation}
\re I_{n_1 n_2}^2(i p_{E0}) = R_{n_1 n_2}(z)\,,\qquad
z = \frac{1}{1+p_{E0}^2}\,,
\label{ReI}
\end{equation}
where
\begin{eqnarray}
R_{n_1 n_2}(z) &=&
\frac{\Gamma^2(n_2-2+\varepsilon)}{\Gamma^2(n_2)}
\left(-\frac{z}{4}\right)^{n_1}
\Biggl[
{}_3\!F_2 \left( \left.
\begin{array}{c}
n_1, \frac{5}{2}-n_2-\varepsilon, 5-n_1-2n_2-2\varepsilon\\
3-n_2-\varepsilon, 5-2n_2-2\varepsilon
\end{array}
\right| z \right)
\nonumber\\
&&{} - 2 A_{n_1 n_2}(z) \cos(\pi\varepsilon)\,
{}_3\!F_2 \left( \left.
\begin{array}{c}
\frac{1}{2}, 3-n_1-n_2-\varepsilon,n_1+n_2-2+\varepsilon\\
3-n_2-\varepsilon,n_2-1+\varepsilon
\end{array}
\right| z \right)
\nonumber\\
&&{} + A_{n_1 n_2}^2(z) \cos(2\pi\varepsilon)\,
{}_3\!F_2 \left( \left.
\begin{array}{c}
1-n_1, n_2-\frac{3}{2}+\varepsilon, n_1+2n_2-4+2\varepsilon\\
n_2-1+\varepsilon, 2n_2-3+2\varepsilon
\end{array}
\right| z \right)
\Biggr]\,,
\label{Rz}
\end{eqnarray}
and
\begin{equation}
A_{n_1 n_2}(z) =
\frac{\pi}{\sin(\pi\varepsilon)}
\frac{\Gamma(n_1+2n_2-4+2\varepsilon)}%
{\Gamma(n_1) h(n_2-2+\varepsilon)}
\left(\frac{z}{4}\right)^{n_2-2+\varepsilon}
\label{Az}
\end{equation}
with $h(x)=x\,\Gamma^2(x)$.
This result simplifies at $n_1=1$:
\begin{eqnarray}
R_{1n_2}(z) &=& - \frac{\Gamma^2(n_2-2+\varepsilon)}{4\Gamma^2(n_2)} z
\Biggl[
{}_3\!F_2 \left( \left.
\begin{array}{c}
1, \frac{5}{2}-n_2-\varepsilon, 4-2n_2-2\varepsilon\\
3-n_2-\varepsilon, 5-2n_2-2\varepsilon
\end{array}
\right| z \right)
\nonumber\\
&&{} - 2 A_{1 n_2}(z) \cos(\pi\varepsilon)\,
{}_2\!F_1 \left( \left.
\begin{array}{c}
\frac{1}{2}, 2-n_2-\varepsilon\\
3-n_2-\varepsilon
\end{array}
\right| z \right)
+ A_{1 n_2}^2(z) \cos(2\pi\varepsilon)
\Biggr]\,.
\label{R1z}
\end{eqnarray}

Calculating the integral
\begin{equation}
I_{n_1 n_2 n_3} =
\frac{\Gamma(n_3-3/2+\varepsilon)}{\pi^{1/2} \Gamma(n_3)}
\int_0^1 R_{n_1 n_2}(z) z^{n_3-3+\varepsilon} (1-z)^{-1/2} dz
\label{I1nm}
\end{equation}
term-by-term, we obtain
\begin{eqnarray}
&&I_{n_1 n_2 n_3} =
\frac{(-1)^{n_1}}{\Gamma^2(n_2)\Gamma(n_3)}
\Biggl[
\frac{h(n_1+n_3-2+\varepsilon)\,
\Gamma^2(n_2-2+\varepsilon) \Gamma(2n_3-3+2\varepsilon)}%
{\Gamma(n_3-1+\varepsilon) \Gamma(2n_1+2n_3-3+2\varepsilon)}
\nonumber\\
&&\qquad{}\times
{}_4\!F_3 \left( \left.
\begin{array}{c}
n_1, \frac{5}{2}-n_2-\varepsilon, n_1+n_3-2+\varepsilon,
5-n_1-2n_2-2\varepsilon\\
n_1+n_3-\frac{3}{2}+\varepsilon, 5-2n_2-2\varepsilon, 3-n_2-\varepsilon
\end{array}
\right| 1 \right)
\nonumber\\
&&{} - 2 \pi \frac{\cos(\pi\varepsilon)}{\sin(\pi\varepsilon)}
\frac{h(n_1+n_2+n_3-4+2\varepsilon)\,
\Gamma(n_1+2n_2-4+2\varepsilon) \Gamma(2n_3-3+2\varepsilon)}%
{(n_2-2+\varepsilon) \Gamma(n_1) \Gamma(n_3-1+\varepsilon)
\Gamma(2n_1+2n_2+2n_3-7+4\varepsilon)}
\nonumber\\
&&\qquad{}\times
{}_4\!F_3 \left( \left.
\begin{array}{c}
\frac{1}{2}, n_1+n_2+n_3-4+2\varepsilon, 3-n_1-n_2-\varepsilon,
n_1+n_2-2+\varepsilon\\
n_1+n_2+n_3-\frac{7}{2}+2\varepsilon, 3-n_2-\varepsilon, n_2-1+\varepsilon
\end{array}
\right| 1 \right)
\nonumber\\
&&{} + \pi^2 \frac{\cos(2\pi\varepsilon)}{\sin^2(\pi\varepsilon)}
\frac{h(n_1+2n_2+n_3-6+3\varepsilon)\,
\Gamma^2(n_1+2n_2-4+2\varepsilon) \Gamma(2n_3-3+2\varepsilon)}%
{\Gamma^2(n_1) \Gamma^2(n_2-1+\varepsilon) \Gamma(n_3-1+\varepsilon)
\Gamma(2n_1+4n_2+2n_3-11+6\varepsilon)}
\nonumber\\
&&\qquad{}\times
{}_4\!F_3 \left( \left.\left.
\begin{array}{c}
1-n_1, n_2-\frac{3}{2}+\varepsilon, n_1+2n_2-4+2\varepsilon,
n_1+2n_2+n_3-6+3\varepsilon\\
n_2-1+\varepsilon, 2n_2-3+2\varepsilon,
n_1+2n_2+n_3-\frac{11}{2}+3\varepsilon
\end{array}
\right| 1 \right)\right].
\label{IRes}
\end{eqnarray}
In particular,
\begin{eqnarray}
&&I_{1 n_2 n_3} = - \frac{1}{\Gamma^2(n_2) \Gamma(n_3)} \Biggl[
\frac{\Gamma^2(n_2-2+\varepsilon) \Gamma(n_3-1+\varepsilon)}%
{2(2n_3-3+2\varepsilon)}
\nonumber\\
&&\qquad{}\times
{}_4\!F_3 \left( \left.
\begin{array}{c}
1, \frac{5}{2}-n_2-\varepsilon, n_3-1+\varepsilon, 4-2n_2-2\varepsilon\\
n_3-\frac{1}{2}+\varepsilon, 5-2n_2-2\varepsilon, 3-n_2-\varepsilon
\end{array}
\right| 1 \right)
\nonumber\\
&&{} - 2 \pi \frac{\cos(\pi\varepsilon)}{\sin(\pi\varepsilon)}
\frac{h(n_2+n_3-3+2\varepsilon) \,
\Gamma(2n_2-3+2\varepsilon) \Gamma(2n_3-3+2\varepsilon)}%
{(n_2-2+\varepsilon) \Gamma(n_3-1+\varepsilon)
\Gamma(2n_2+2n_3-5+4\varepsilon)}
\nonumber\\
&&\qquad{}\times
{}_3\!F_2 \left( \left.
\begin{array}{c}
\frac{1}{2}, n_2+n_3-3+2\varepsilon, 2-n_2-\varepsilon\\
n_2+n_3-\frac{5}{2}+2\varepsilon, 3-n_2-\varepsilon
\end{array}
\right| 1 \right)
\nonumber\\
&&{} + \pi^2 \frac{\cos(2\pi\varepsilon)}{\sin^2(\pi\varepsilon)}
\frac{h(2n_2+n_3-5+3\varepsilon) \,
\Gamma^2(2n_2-3+2\varepsilon) \Gamma(2n_3-3+2\varepsilon)}%
{\Gamma^2(n_2-1+\varepsilon) \Gamma(n_3-1+\varepsilon)
\Gamma(4n_2+2n_3-9+6\varepsilon)}
\Biggr]\,.\label{I1Res}
\end{eqnarray}

For example, let us consider the convergent integral $I_{122}$.
Using the reduction procedure of Ref.~\cite{GSS:06},
we can relate it to the master integral $I_2\equiv I_{111}$:
\begin{equation}
I_{122} = -
\frac{(d-3)^2 (d-4) (3d-8) (3d-10)}{8 (3d-11) (3d-13)} I_2\,.
\label{I2}
\end{equation}
From Eq.~(\ref{I1Res}) we have
\begin{eqnarray}
\frac{I_{122}}{\Gamma^3(1+\varepsilon)} &=&
- \frac{1}{2\varepsilon^2} \Biggl[ \frac{1}{1+2\varepsilon}
{}_4\!F_3 \left( \left.
\begin{array}{c}
1, \frac{1}{2}-\varepsilon, 1+\varepsilon, -2\varepsilon\\
\frac{3}{2}+\varepsilon, 1-\varepsilon, 1-2\varepsilon
\end{array}
\right| 1 \right)
\nonumber\\
&&{} - \frac{2}{1+4\varepsilon}
\frac{\Gamma^2(1-\varepsilon) \Gamma^3(1+2\varepsilon)}%
{\Gamma^2(1+\varepsilon) \Gamma(1-2\varepsilon) \Gamma(1+4\varepsilon)}
{}_3\!F_2 \left( \left.
\begin{array}{c}
\frac{1}{2}, 1+2\varepsilon, -\varepsilon\\
\frac{3}{2}+2\varepsilon, 1-\varepsilon
\end{array}
\right| 1 \right)
\nonumber\\
&&{} + \frac{1}{1+6\varepsilon}
\frac{\Gamma^2(1-\varepsilon) \Gamma^4(1+2\varepsilon)
\Gamma(1-2\varepsilon) \Gamma^2(1+3\varepsilon)}%
{\Gamma^4(1+\varepsilon) \Gamma(1+4\varepsilon)
\Gamma(1-4\varepsilon) \Gamma(1+6\varepsilon)}
\Biggr]\,.
\label{I122}
\end{eqnarray}

There are several methods to expand the hypergeometric functions
in $\varepsilon$~\cite{Kalmykov,W:04,hypexpII}.
We first follow the method of Ref.~\cite{W:04}.
First we express $\Gamma$ functions in the hypergeometric series
as exponents containing $S$-sums
\begin{equation}
S(N;k_1,\ldots,k_l;x_1,\ldots,x_l) =
\sum_{N\geqslant n_1\geqslant\cdots\geqslant n_l\geqslant 1}
\frac{x_1^{n_1}}{n_1^{k_1}} \cdots \frac{x_l^{n_l}}{n_l^{k_l}}\,.
\label{nested}
\end{equation}
Some indices are integer at $\varepsilon\to0$ and some are half-integer;
therefore, we get both $S(n;k;1)$ and $S(2n;k;1)$.
We can re-express $S(n;k;1)$ via $S(2n;k;\pm1)$
by inserting $(1+(-1)^{n_1})/2$ under the summation sign:
\begin{eqnarray}
&&{}_3\!F_2 \left( \left.
\begin{array}{c}
\frac{1}{2}, 1+2\varepsilon, -\varepsilon\\
\frac{3}{2}+2\varepsilon, 1-\varepsilon
\end{array}
\right| 1 \right)
= 1
\nonumber\\
&&{} - \varepsilon (1+4\varepsilon) \sum_{n=1}^\infty
\frac{1}{(n-\varepsilon)(2n+1+4\varepsilon)}
\exp\left[ - \sum_{k=1}^\infty \frac{(-4\varepsilon)^k}{k}
S(2n;k;-1) \right]\,,
\nonumber\\
&&{}_4\!F_3 \left( \left.
\begin{array}{c}
1, \frac{1}{2}-\varepsilon, 1+\varepsilon, -2\varepsilon\\
\frac{3}{2}+\varepsilon, 1-\varepsilon, 1-2\varepsilon
\end{array}
\right| 1 \right)
= 1
\nonumber\\
&&{} - 2 \varepsilon (1+2\varepsilon) \sum_{n=1}^\infty
\frac{1}{(n-2\varepsilon)(2n+1+2\varepsilon)}
\exp\left[ \sum_{k=1}^\infty \frac{(2\varepsilon)^k}{k}
\left(1 - (-1)^k\right)
S(2n;k;-1) \right] ,
\label{hyper}
\end{eqnarray}
Products of $S$-sums with upper limit $2n$
are expressed in terms of single $S$-sums
by means of the well-known algebra~\cite{Broadhurst,V:99,MUW:02}.
We expand the rational factors in Eq.~(\ref{hyper}) in $\varepsilon$,
and then expand them into partial fractions.
After that, separate sums may diverge,
and we introduce an upper limit $N$ instead of $\infty$ in the outermost sum.
Sums with $1/n^k$ can be re-written as sums in $n$ to $2N$
by inserting $(1+(-1)^n)/2$, and those with $1/(2n+1)^k$ ---
by inserting $(1-(-1)^n)/2$:
\begin{eqnarray*}
&&\sum_{n=1}^{N} \frac{1}{n^k} S(2n;k_1,\ldots,k_l;x_1,\ldots,x_l)\\
&&{} = 2^{k-1} \bigl[S(2N;k,k_1,\ldots,k_l;1,x_1,\ldots,x_l)
+ S(2N;k,k_1,\ldots,k_l;-1,x_1,\ldots,x_l)\bigr]\,,\\
&&\sum_{n=1}^{N} \frac{1}{(2n+1)^k} S(2n;k_1,\ldots,k_l;x_1,\ldots,x_l)\\
&&{} = \frac{1}{2} \bigl[S(2N;k,k_1,\ldots,k_l;1,x_1,\ldots,x_l)
- S(2N;k,k_1,\ldots,k_l;-1,x_1,\ldots,x_l)\\
&&\hphantom{{}=\frac{1}{2}\bigl[\bigr.}
- S(2N;k+k_1,k_2,\ldots,k_l;x_1,x_2,\ldots,x_l)
+ S(2N;k+k_1,k_2,\ldots,k_l;-x_1,x_2,\ldots,x_l)\bigr]
\end{eqnarray*}
(terms vanishing at $N\to\infty$ are omitted here).
After that, all sums divergent at $N\to\infty$ cancel;
in the remaining sums, we may set $N=\infty$.
They are related to the Euler--Zagier sums~\cite{Broadhurst,V:99}.

Yet another method of expanding certain classes of
hypergeometric functions about half-integer parameters
is provided by the algorithm of Ref.~\cite{hypexpII},
by means of which we have calculated the expansion of Eq.~(\ref{I122})
up to order ${\cal O}(\varepsilon^7)$.
After factoring out an appropriate combination of pre-factors,
the expansion reads
\begin{eqnarray}
\frac{I_{122}}{\Gamma^3(1+\varepsilon)} &=&\frac{\pi^2}{3(1+6\varepsilon)}
\nonumber\\
&&\times\biggl[1 - \pi^2 \varepsilon^2 + 48\zeta_3 \varepsilon^3
- \frac{38\pi^4}{15}\varepsilon^4
- 48 \left(\pi^2\zeta_3-30 \zeta_5\right)\varepsilon^5
+ \left(1152 \zeta_3^2-\frac{4793\pi^6}{945}\right)\varepsilon^6
\nonumber\\
&&{} + \left(39312 \zeta_7-\frac{608\pi^4\zeta_3}{5}
-1440\pi^2\zeta_5\right)\varepsilon^7
+\mathcal{O}(\varepsilon^8) \biggr]\,.
\label{eq:exp}
\end{eqnarray}
In order to check the correctness of this expansion,
we have converted the hypergeometric functions in Eq.~(\ref{I122})
to single Mellin-Barnes representations
and subsequently obtained the coefficients numerically~\cite{C:05}.
We find agreement to at least 14 decimal digits.

It is now straightforward to obtain the expansion
of the master integral $I_2$ from Eq.~(\ref{I2}):
\begin{equation}
\frac{I_2}{\Gamma^3(1+\varepsilon)} =
- \frac{\pi^2}{6} \biggl[ \frac{1}{\varepsilon} + \frac{5}{2}
- \left( \pi^2 + \frac{21}{4} \right) \varepsilon
+ \left( 48 \zeta_3 - \frac{5}{2} \pi^2 - \frac{599}{8} \right) \varepsilon^2
+ \cdots \biggr]\,.
\label{I2res}
\end{equation}
The first two terms have been obtained in Ref.~\cite{GSS:06}
by a completely different method.
The next one has been found by A.~Pak from the requirement of
cancellation of $1/\varepsilon$ poles in a physical calculation.
The last term is then required to calculate the corresponding finite part.
Yet higher terms in the expansion of $I_2$ can be obtained
from Eq.~(\ref{eq:exp}).

As a closing remark we would like to mention that the expansion
in Eq.~(\ref{eq:exp}) is found to agree up to order ${\cal O}(\varepsilon^7)$
with the expansion of the simple formula
\begin{equation}
\frac{I_{122}}{\Gamma^3(1+\varepsilon)} =
\frac{\pi^2}{3}
\frac{\Gamma^3(1+2\varepsilon) \Gamma^2(1+3\varepsilon)}%
{\Gamma^6(1+\varepsilon) \Gamma(2+6\varepsilon)}\,.
\label{danielmaitrefussballgott}
\end{equation}
We do not know any analytical proof of this result,
but high-precision numerical tests of Eqs.~(\ref{I122})
and~(\ref{danielmaitrefussballgott}) for various values
of $\varepsilon$ on the real axis and in the complex plane
strengthen the conjecture that this result might indeed be valid to all orders.
However, until an analytical proof of Eq.~(\ref{danielmaitrefussballgott})
is found, we are guaranteed to obtain the correct expansion
only by the analytically derived expression~(\ref{I122}).

\acknowledgments
A.G.\ is grateful to O.V.~Tarasov for the suggestion to use a method
similar to Ref.~\cite{B:92};
to A.~Pak for communicating his result for the $\mathcal{O}(\varepsilon)$ term
in Eq.~(\ref{I2res}) and for motivation to calculate
the $\mathcal{O}(\varepsilon^2)$ term;
to A.I.~Davydychev for an advice on hypergeometric functions;
and to V.A.~Smirnov for an independent check of some calculations.
T.H.\ is supported by Deutsche Forschungsgemeinschaft,
SFB/TR 9 ``Computergest\"{u}tzte Theoretische Teilchenphysik''.
D.M.\ is supported by the SNF under contract PBZH2-117028
and by the US Departement of Energy under contract
DE-AC02-76SF00515.

\end{document}